\documentstyle[twoside,fleqn,espcrc2,epsf]{article}

\newcommand{\half}{\mbox{\small $\frac{1}{2}$}}          
\newcommand{\third}{\mbox{\small $\frac{1}{3}$}}         
\newcommand{\twothird}{\mbox{\small $\frac{2}{3}$}}      

\def\lsim{\mathrel{\rlap{\lower4pt\hbox{\hskip1pt$\sim$}}
    \raise1pt\hbox{$<$}}}                
\def\gsim{\mathrel{\rlap{\lower4pt\hbox{\hskip1pt$\sim$}}
    \raise1pt\hbox{$>$}}}                

\def\3{\ss}

\newcommand{\AmS}{{\protect\the\textfont2
  A\kern-.1667em\lower.5ex\hbox{M}\kern-.125emS}}

\title{
       \vspace{-3.65cm}                                     
       {\normalsize DESY 98--124}    \\[-0.2cm]             
       {\normalsize HUB--EP--98/46}  \\[-0.2cm]             
       {\normalsize FUB-HEP/3-98}    \\[-0.2cm]             
       {\normalsize TPR-98-29}       \\[-0.2cm]             
       {\normalsize September 1998}  \\                     
       \vspace{1.32cm}                                      
       Nucleon form factors and $O(a)$ Improvement%
            \thanks{Talk given by R. Horsley at Lat98,      
                    Boulder, U.S.A.}}                       

\author{S.~Capitani%
           \address{Deutsches Elektronen-Synchrotron DESY,
                    D-22603 Hamburg, Germany},
        M.~G\"ockeler%
           \address{Institut f\"ur Theoretische Physik, Universit\"at
                    Regensburg, D-93040 Regensburg, Germany},
        R.~Horsley%
           \address{Institut f\"ur Physik, Humboldt-Universit\"at zu Berlin,
                    D-10115 Berlin, Germany},
        B.~Klaus%
           \address{Institut f\"ur Theoretische Physik,
                    Freie Universit\"at Berlin, D-14195 Berlin, Germany},
        H.~Oelrich%
           \address{Deutsches Elektronen-Synchrotron DESY and NIC,
                    D-15735 Zeuthen, Germany},
        H.~Perlt%
           \address{Institut f\"ur Theoretische Physik, Universit\"at
                    Leipzig, D-04109 Leipzig, Germany},
        D.~Petters$^{\rm d,}$ \hspace{-0.2cm} $^{\rm e}$,
        D.~Pleiter$^{\rm d,}$ \hspace{-0.2cm} $^{\rm e}$,
        P.~E.~L. Rakow$^{\rm b}$,
        G.~Schierholz$^{\rm a,}$ \hspace{-0.2cm} $^{\rm e}$,
        A.~Schiller$^{\rm f}$
        and
        P.~Stephenson%
           \address{Dipartimento di Fisica,
                    Universit\`a degli Studi di Pisa e INFN,
                    Sezione di Pisa, 56100 Pisa, Italy}}
       
\begin{document}

\begin{abstract}
Nucleon form factors have been extensively
studied both experimentally and theoretically for many years.
We report here on new results of a high statistics quenched
lattice QCD calculation of vector and axial-vector
nucleon form factors at low momentum transfer within the
Symanzik improvement programme. The simulations are performed
at three $\kappa$ and three $\beta$ values allowing first an extrapolation
to the chiral limit and then an extrapolation in the lattice spacing to the
continuum limit. The computations are all fully non-perturbative.
A comparison with experimental results is made.
\end{abstract}

\maketitle

\setcounter{footnote}{0}


\section{INTRODUCTION}
\label{intro}

For many years experiments have been performed with electron--nucleon
scattering to obtain information about the structure of the nucleon.
Form factors are defined from the general decomposition of the proton, $p$
(or neutron, $n$) matrix element%
\footnote{We have already re-written everything in euclidean space,
so that eg $p=(iE_p,\vec{p})$ and $-q^2 \equiv q^{({\cal M})2} > 0$.}
($q = p - p^\prime$):
\begin{eqnarray}
  \lefteqn{\langle \vec{p},\vec{s}|
           \widehat{\cal V}^{\twothird u - \third d}_\mu(\vec{q})
                          |\vec{p}^\prime,\vec{s}^\prime \rangle = }
         &                                    \nonumber \\
         &  \overline{u}(\vec{p},\vec{s})
              \left[ \gamma_\mu 
                     F^p_1 +
                     \sigma_{\mu\nu}
                     {q_\nu \over 2m} F^p_2
              \right] u(\vec{p}^\prime,\vec{s}^\prime).
                                                \nonumber
\end{eqnarray}
We have $F_1(0) = 1$ as ${\cal V}$ is a conserved
current, while $F_2(0) = \mu - 1$ measures the anomalous magnetic
moment (in magnetons). Usually we define the Sachs form factors:
\begin{eqnarray}
   G_e(-q^2) &=& F_1(-q^2) + {-q^2\over (2m)^2} F_2(-q^2),
                                                \nonumber \\
   G_m(-q^2) &=& F_1(-q^2) + F_2(-q^2).
                                                \nonumber
\end{eqnarray}
Experiments lead to phenomenological dipole fits:
\begin{eqnarray}
   G_e^p(-q^2) &\sim& {G_m^p(-q^2) \over \mu^p}
                              \quad \sim \quad
                              {G_m^n(-q^2) \over \mu^n}
                                                \nonumber \\  
                          &=&
                            1 /
                            \left(1+ \left( -q^2 / m_V^2 \right)
                            \right)^2, 
                                                \nonumber \\  
   G_e^n(-q^2) &\sim& 0,
                                                \nonumber
\end{eqnarray}
with $m_V \sim 0.82$ $\mbox{GeV}$, $\mu^p \sim 2.79$,
$\mu^n \sim -1.91$.

Neutrino--neutron scattering, $n\nu_\mu \to p\mu^-$,
gives from the charged weak current the axial form factor $g_A(-q^2)$. 
In addition $g_A(0)$ is also accurately obtained from $\beta$-decay,
$n \to pe^-\overline{\nu}$. Upon using current algebra this
form factor can be related to the matrix element:
\begin{eqnarray}
  \lefteqn{\langle \vec{p},\vec{s}| \widehat{\cal A}^{u-d}_\mu(\vec{q})
                          |\vec{p}^\prime,\vec{s}^\prime \rangle = }
         &                                    \nonumber \\
         &  \overline{u}(\vec{p},\vec{s})
              \left[ \gamma_\mu\gamma_5
                    g_A + i\gamma_5 {q_\mu \over 2m} h_A
              \right] u(\vec{p}^\prime,\vec{s}^\prime).
                                                \nonumber
\end{eqnarray}
The phenomenological fits are:
\begin{eqnarray}
   g_A(q^2) =  g_A(0) /
                       \left(1+ \left( -q^2 / m_A^2 \right)
                       \right)^2,
                                                \nonumber
\end{eqnarray}
with $g_A(0) = 1.26$, $m_A \sim 1.00$ $\mbox{GeV}$.

\section{THE LATTICE METHOD}
\label{method}

\begin{figure}[ht]
   \epsfxsize=6.50cm \epsfbox{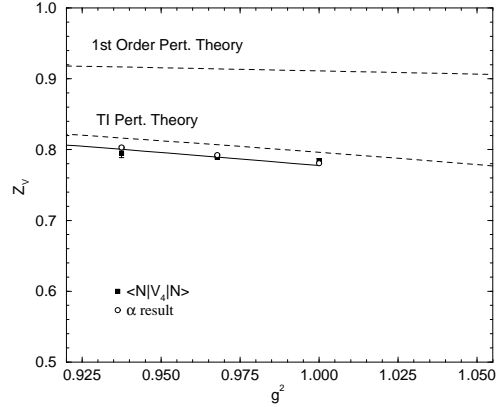}
   \vspace*{-1.00cm}
   \caption{\footnotesize 
            $Z_V$ for improved fermions. Shown is the lowest order
            perturbation result together with a tadpole-improved
            version (as given in \cite{goeckeler97a}).
            The non-perturbative determinations are shown as
            open circles, \cite{luescher97a}, and filled squares,
            this work.}
   \vspace*{-0.75cm}
   \label{fig_zv_g2_magic}
\end{figure}

Quenched configurations have been generated at
$\beta=6.0$ ($O(500)$, $16^3\times 32$ lattice)
$\beta=6.2$ ($O(300)$, $24^3\times 48$ lattice) and
$\beta=6.4$ ($O(100)$, $32^3\times 48$ lattice), \cite{pleiter98a}.
By forming the ratio of three-to-two point functions, \cite{martinelli89a}:
\begin{eqnarray}
   \lefteqn{R_{\alpha\beta}(t,\tau; \vec{p},\vec{q})
      = {\langle N_\alpha(t;\vec{p}) {\cal O}(\tau;\vec{q})
        \overline{N}_\beta(0;\vec{p}^\prime) \rangle \over
        \langle N(t;\vec{p}) \overline{N}(0;\vec{p}) \rangle} \times}
     &                                     \nonumber \\  
     &  \hspace{-0.25cm} \left[
         {\langle N(\tau;\vec{p}) \overline{N}(0;\vec{p}) \rangle
          \langle N(t;\vec{p}) \overline{N}(0;\vec{p}) \rangle
          \langle N(t-\tau;\vec{p}^\prime)
                  \overline{N}(0;\vec{p}^\prime) \rangle
         \over
         \langle N(\tau;\vec{p}^\prime)
               \overline{N}(0;\vec{p}^\prime) \rangle
         \langle N(t;\vec{p}^\prime)
         \overline{N}(0;\vec{p}^\prime) \rangle
         \langle N(t-\tau;\vec{p})
               \overline{N}(0;\vec{p}) \rangle} \right]^{\half}
                                       \nonumber \\  
     &                                 \nonumber \\  
     & \propto \langle N_\alpha(\vec{p})| \widehat{\cal O}(\vec{q})
                                    |N_\beta(\vec{p}^\prime) \rangle,
                                                \nonumber
\end{eqnarray}
the appropriate matrix elements can be found. (Only the
quark line connected part of the $3$-point function is considered.)
For each $\beta$ we chose three $\kappa$ values
and a variety of $3$-momenta:
$\vec{p} = 2\pi/N_s \{$ $(0,0,0)$, $(1,0,0)$ $\}$,
$\vec{q} = 2\pi/N_s \{$ $(0,0,0)$, $(0,1,0)$, $(0,2,0)$, $(1,0,0)$,
$(2,0,0)$, $(1,1,0)$, $(1,1,1)$, $(0,0,1)$ $\}$ together
with the nucleon either unpolarised or polarised in the $y$ direction. 
(Some combinations were too noisy to be used though.)
After sorting the matrix elements into $q^2$ classes (defined by $q^2$
in the chiral limit), $4$-parameter fits are made assuming 
that the form factors are linear in the bare quark mass $am_q$.
$O(a)$ improved Symanzik operators are used:
\begin{eqnarray}
   {\cal V}_\mu^R &=& Z_V ( 1 + b_Vam_q ) \times
                                                \nonumber \\  
                  & & \quad \left[ \bar{\psi}\gamma_\mu\psi
                            + \half i c_V
                    a\partial_\lambda(\bar{\psi}\sigma_{\mu\lambda}\psi)
                            \right],
                                                \nonumber \\  
   {\cal A}_\mu^R &=& Z_A ( 1 + b_Aam_q ) \times
                                                \nonumber \\  
                  & & \quad  \left[ \bar{\psi}\gamma_\mu\gamma_5\psi
                      + \half c_Aa\partial_\mu(\bar{\psi}\gamma_5\psi)
                              \right],
                                                \nonumber
\end{eqnarray}
where $Z_V$, $Z_A$, $b_V$, $c_V$, $c_A$ (and $c_{sw}$)
have been non-perturbatively calculated by the Alpha
collaboration, \cite{luescher97a}.
All matrix elements thus are correct to $O(a^2)$. We can check
$Z_V$ as ${\cal V}_\mu$ is a conserved current (ie $F_1(0)=1$).
In Fig.~\ref{fig_zv_g2_magic} 
we show a comparison of the two determinations of $Z_V$. Very good
agreement is seen. This is not the case when Wilson fermions are used
(see ref.~\cite{capitani98a}). Finally we note
that although we have included the improvement terms in our operators,
numerically they seem to have little influence on the value
of the matrix element.

\section{RESULTS}
\label{results}

\begin{figure}[t]
   \vspace*{-0.25cm}
   \hspace*{0.25cm}
   \epsfxsize=6.25cm \epsfbox{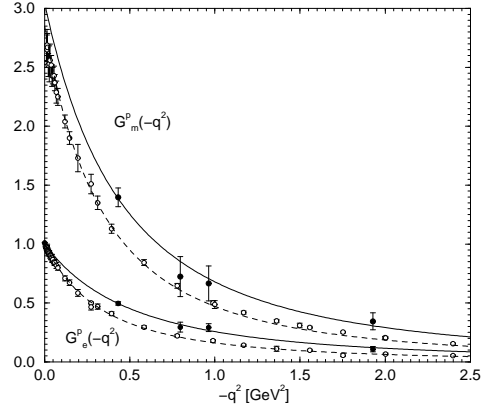}
   \vspace*{-1.00cm}
   \caption{\footnotesize 
            The proton form-factors $G^p_e(-q^2)$ and $G^p_m(-q^2)$
            against $-q^2$ showing experimental results (open circles,
            taken from ref.~\cite{christov96a})
            and lattice results (filled circles, $\beta=6.2$ only).
            The string tension is used to fix the
            scale as in \cite{goeckeler97a}. All fits are dipole fits.}
   \vspace*{-0.75cm}
   \label{fig_Ge+Gm_Vff_Qu-Qd_GeV2_b6p20}
\end{figure}

\begin{figure}[ht]
   \vspace*{-0.50cm}
   \epsfxsize=6.50cm \epsfbox{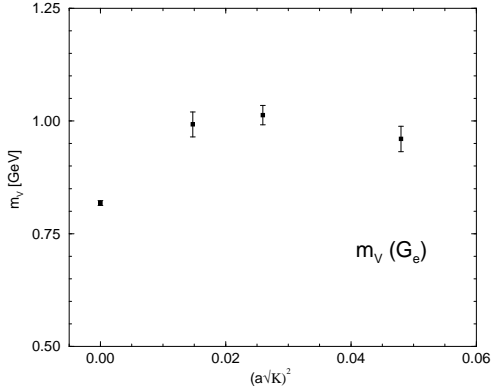} 
   \vspace*{-1.00cm}
   \caption{\footnotesize 
            $m_V$ from $\beta=6.0$, $6.2$, $6.4$ against $a^2$.
            The phenomenological value is also given at $a^2=0$.}
   \vspace*{-0.75cm}
   \label{fig_Ge_p.fitparms}
\end{figure}

\begin{figure}[ht]
   \hspace*{0.25cm}
   \epsfxsize=6.00cm \epsfbox{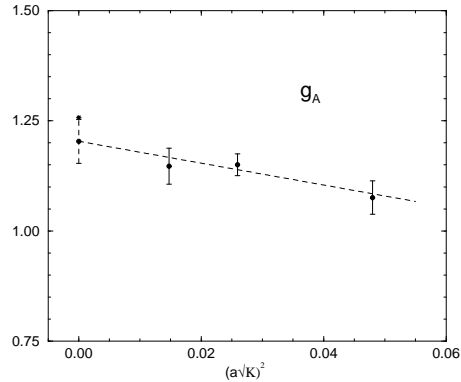}        
   \vspace*{-1.00cm}
   \caption{\footnotesize 
            The continuum extrapolation of $g_A(0)$.}
   \vspace*{-0.75cm}
   \label{fig_ga_sig2}
\end{figure}

In Fig.~\ref{fig_Ge+Gm_Vff_Qu-Qd_GeV2_b6p20}
we show $G^p_e(-q^2)$ and $G^p_m(-q^2)$ for $\beta=6.2$ together
with experimental results (also plotting the other $\beta$ values tends
to clutter the picture). Making dipole fits gives
Fig.~\ref{fig_Ge_p.fitparms} for the continuum extrapolation.
There seems to be little inclination for $m_V$ to approach
the experimental result.
(A roughly similar result is obtained from $G^p_m$, although due 
to larger error bars the results are more compatible.)
\begin{figure}[ht]
   \vspace*{-0.20cm}
   \hspace*{0.25cm}
   \epsfxsize=5.875cm \epsfbox{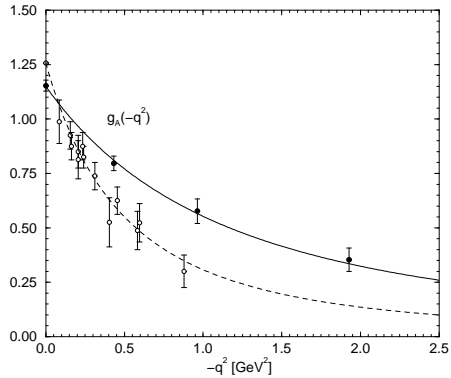}
   \vspace*{-1.00cm}
   \caption{\footnotesize 
            $g_A(-q^2)$ against $-q^2$,
            notation as in Fig.~\ref{fig_Ge+Gm_Vff_Qu-Qd_GeV2_b6p20}.}
   \vspace*{-0.75cm}
   \label{fig_GA_Adr_1u-1d_GeV2_b6p20}
\end{figure}
\begin{figure}[ht]
   \vspace*{-0.50cm}
   \epsfxsize=6.50cm \epsfbox{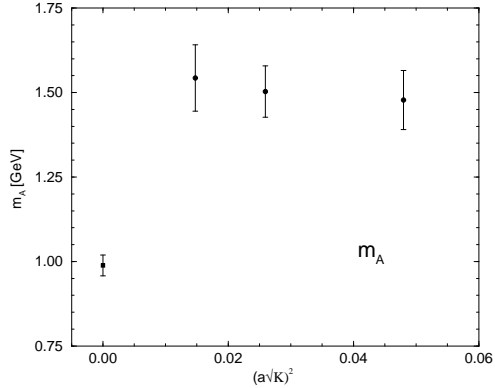}  
   \vspace*{-1.00cm}
   \caption{\footnotesize 
            The continuum extrapolation of $m_A$.}
   \vspace*{-0.75cm}
   \label{fig_GA_p.fitparms}
\end{figure}
For the axial current we find the results in 
Figs.~\ref{fig_GA_Adr_1u-1d_GeV2_b6p20},~\ref{fig_GA_p.fitparms}.
The form factor fall-off is again too soft as $m_A$ is too large.
However the important $g_A(0)$ is faring better, see Fig.~\ref{fig_ga_sig2}.

\section{CONCLUSIONS}
\label{conclusions}

We have performed simulations at three $\beta$ values so that an
attempt can be made to take the continuum extrapolation, $a \to 0$.
While the lattice dipole masses seem to be too large,
$g_A(0)$ is in reasonable agreement with the experimental result.
The mass discrepancies may be due to a quenching effect, although
only similar simulations using dynamical fermions will
be able to answer this.

\section*{ACKNOWLEDGEMENTS}
\label{acknowledgement}

The numerical calculations were performed on the
Ape $QH2$ at DESY-Zeuthen and the Cray $T3E$ at ZIB, Berlin.
Financial support from the DFG is also gratefully acknowledged.


\begin{thebibliography}{9}
\bibitem{pleiter98a}     D. Pleiter, this conference.
\bibitem{martinelli89a}  G. Martinelli et al., Nucl. Phys. B316 (1989) 355;
                         W. Wilcox et al., Phys. Rev. D46 (1992) 1109,  
                         hep-lat/9205015 ; K.~F. Liu et al., Phys. Rev.
                         D49 (1994) 4755, hep-lat/9305025.
\bibitem{luescher97a}    M. L\"uscher et al.,
                         Nucl. Phys. B491 (1997) 323, hep-lat/9609035;
                         Nucl. Phys. B491 (1997) 344, hep-lat/9611015;
                         M. Guagnelli et al., Nucl. Phys.
                         (Proc Suppl) 63 (1998) 886, hep-lat/9709088.
\bibitem{goeckeler97a}   M. G\"ockeler et al., Phys. Rev. D57 (1998) 5562,
                         hep-lat/9707021.
\bibitem{capitani98a}    S. Capitani et al., $31^{st}$ Ahrenshoop
                         Symposium, Buckow, Germany (Wiley-VCH 1998),
                         hep-lat/9801034.
\bibitem{christov96a}    C.~V. Christov et al., Prog. Part. Nucl. Phys. 37
                         (1996), hep-ph/9604441.
\end{thebibliography}
\end{document}